\documentclass[aps,twocolumn,showpacs,preprintnumbers,amsmath,amssymb,superscriptaddress]{revtex4-1}

\usepackage{graphicx}
\usepackage{epstopdf}
\usepackage{amsmath}
\usepackage{bm}
\usepackage{xcolor,xspace}
\usepackage[ulem=normalem]{changes}
\usepackage{hyperref}
\usepackage{xfrac}
\hypersetup{colorlinks,linkcolor={red},citecolor={blue},urlcolor={blue}}

\definechangesauthor[name=Maurycy, color=blue]{1}

\begin{document}
\title{Shell-model study of octupole collectivity near $^{208}$Pb}
\author{P.~Van~Isacker}
\affiliation{Grand Acc\'el\'erateur National d'Ions Lourds, CEA/DRF-CNRS/IN2P3,
Boulevard Henri Becquerel, F-14076 Caen, France}
\author{M.~Rejmund}
\affiliation{Grand Acc\'el\'erateur National d'Ions Lourds, CEA/DRF-CNRS/IN2P3,
Boulevard Henri Becquerel, F-14076 Caen, France}

\date{\today}

\begin{abstract} 
We show that the collectivity of the particle-hole wave function
of low-lying octupole $3^-$ states in doubly magic nuclei
is mainly due to the neutron-proton interaction.
Both the enhanced reduced transition probability to the ground state,
$B({\rm E3};3^-_1\rightarrow0^+_1)$,
and the coupling of the octupole excitation to a nucleon
result from the coherent action of all the components of the collective state.
The results obtained with a realistic shell-model interaction both for $^{208}$Pb and $^{209}$Pb
agree with the geometric collective model of Bohr and Mottelson,
where octupole excitations are associated with phonons
corresponding to collective shape oscillations of the surface of the nucleus.
\end{abstract}

\maketitle

{\it Introduction.}
Collective excitations are ubiquitous in quantum many-body systems. 
Their occurrence has given rise to models
that capture essential properties in terms of a few degrees of freedom,
usually of bosonic nature.
Ultimately, however, their description invokes excitations of the constituent particles of the system,
which may vary from fully collective to purely single-particle.
Nuclei are prime examples of this dichotomy.
Depending on its location in the nuclear chart
(i.e., its number of neutrons $N$ and protons $Z$),
a nucleus may exhibit either single-particle or collective types of excitation.
To complicate matters even further, in many nuclei both types of excitation coexist.
This dual nature is also reflected in the history of nuclear physics:
the 1950s saw the development of the nuclear shell model~\cite{Mayer49,Haxel49},
which stresses the importance of single-particle behavior,
in parallel with the elaboration of the geometric collective model~\cite{Rainwater50,Bohr53}.
It is by now understood that both descriptions of the nucleus are not necessarily incompatible;
for example, nuclear rotational motion arises as a result of an SU(3) symmetry of the shell model~\cite{Elliott58a,Elliott58b}.
Nevertheless, the interplay between single-particle and collective nuclear excitations
to this day remains to be fully understood microscopically.

In many doubly magic nuclei the lowest-energy excitation has spin-parity $J^\pi=3^-$.
In the geometric collective model this state is described
as a surface vibration in the octupole degree of freedom.
In the shell model it corresponds to a {\em coherent} superposition
of particle-hole (ph) excitations of nucleons across the shell closures
and we show here that this superposition obeys universal symmetry properties.
Nuclei with one nucleon more or one nucleon less than the double shell closure
are then expected to exhibit excitations associated with the single particle or hole
as well as those where the single nucleon is coupled to the octupole vibration~\cite{BM75}. 
Such odd-mass nuclei are therefore ideal testing grounds of the interplay
between single-particle and collective nuclear excitations.

{\it Octupole states in doubly magic nuclei.}
In a one-particle one-hole (1p1h) approximation an octupole excitation in a doubly magic nucleus
corresponds to the linear combination of 1p1h excitations across the shell closure 
\begin{equation}
|3^-_{\rm c}\rangle=
\sum_\rho\sum_{k'k}c^\rho_{k'k}|j_{\rho k'}j_{\rho k}^{-1};3^-\rangle
\label{e_oph}
\end{equation}
where particle orbitals above the shell closure
are denoted with primed indices  $j_{\rho k'}$
[short-hand notation for the complete set of single-particle labels $(nlj)_{\rho k'}$]
and hole orbitals below the shell closure
with unprimed indices $j_{\rho k}$.
Both occur for neutrons ($\rho=\nu$) as well as protons ($\rho=\pi$).
The coefficients $c^\rho_{k'k}$ in Eq.~(\ref{e_oph})
are the amplitudes of the wave function,
with $(c^\rho_{k'k})^2$ being the probability
to find the $|3^-_{\rm c}\rangle$ state in the $|j_{\rho k'}j_{\rho k}^{-1};3^-\rangle$ configuration.
The values of  $c^\rho_{k'k}$ 
result from the diagonalization of the nuclear Hamiltonian
\begin{equation}
\hat H=
\sum_\rho\biggl(\sum_k\epsilon_{\rho k}\hat n_{\rho k}+\sum_{k'}\epsilon_{\rho k'}\hat n_{\rho k'}+
\hat V_{\rho\rho}\biggr)+\hat V_{\nu\pi},
\label{e_ham}
\end{equation}
where $\hat n_{\rho k}$ is the number operator for $\rho$ nucleons in orbital $j_{\rho k}$,
$\epsilon_{\rho k}$ is its single-nucleon energy,
and $\hat V_{\nu\nu}$, $\hat V_{\pi\pi}$, and $\hat V_{\nu\pi}$
are the neutron-neutron ($\nu\nu$), proton-proton ($\pi\pi$),
and neutron-proton ($\nu\pi$) interactions, respectively.
The orbitals appropriate for the $^{208}$Pb region span two major shells for neutrons and protons,
from $N=82$ to $184$ for neutrons and from $Z=50$ to $126$ for protons.
The effective single-nucleon energies have been deduced
from the experimental data by Warburton and Brown~\cite{Warburton1991}. 
There are $\sim$35000 two-body matrix elements in this space,
obtained in a variety of ways, as described by Brown~\cite{Brown00}. 
The set used here is taken from Ref.~\cite{Wrzesinski01}.

\begin{figure}
\includegraphics[width=8.5cm]{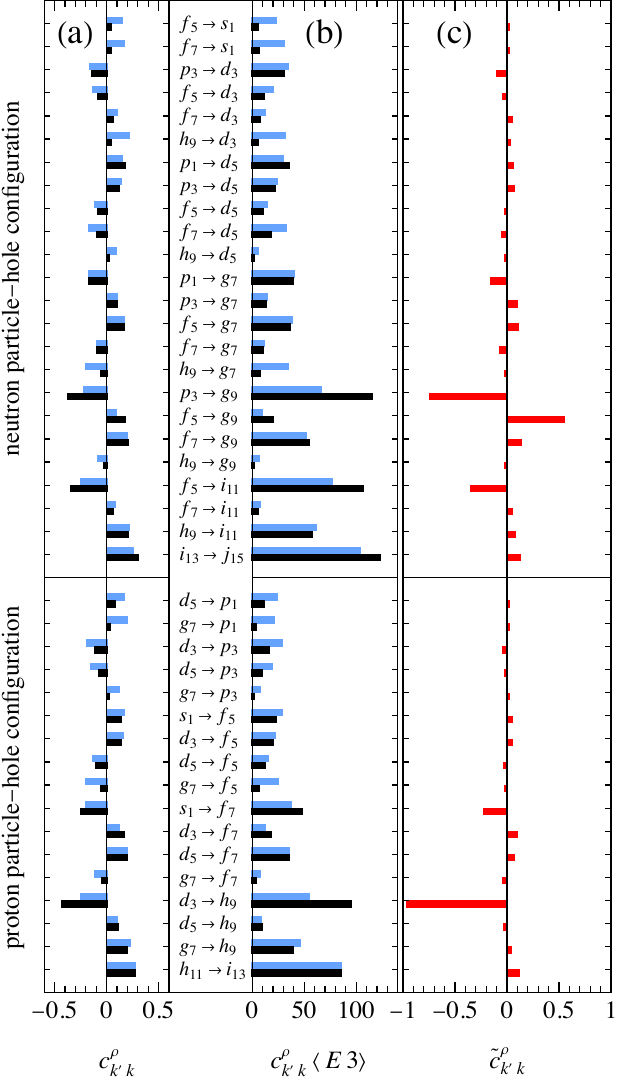}
\caption{(a) Components in the wave function of the collective $3^-_{\rm c}$ state
calculated with a realistic shell-model Hamiltonian ($c^\rho_{k'k}$, black)
and compared with the analytic expression~(\ref{e_can})
($\bar c^\rho_{k'k}$ with $\alpha_\nu=0.734$ and $\alpha_\pi=0.679$, blue).
(b) The reduced transition matrix elements 
$\langle{\rm E3}\rangle\equiv\langle0^+_1\|{\rm E3}\|j_{\rho k'}j_{\rho k}^{-1};3^-\rangle$,
in units $e$fm$^3$,
multiplied with the coefficients $c^\rho_{k'k}$ (black) and $\bar c^\rho_{k'k}$ (blue).
(c) Components calculated with
a realistic shell-model Hamiltonian with
zero $\nu\pi$ interaction ($\tilde c^\rho_{k'k}$, red).
On the $y$ axis are indicated all possible 1p1h configurations in a simplified notation,
e.g.\ $f_5\rightarrow s_1$ stands for $|s_{1/2}f_{5/2}^{-1};3^-\rangle$.}
\label{f_wav3} 
\end{figure}
With this realistic shell-model Hamiltonian
the first-excited $3^-$ level in $^{208}$Pb is calculated at an excitation energy $E_{\rm x}=2460$~keV,
to be compared with the experimental value of 2615~keV~\cite{NNDC}.
The state has a highly fragmented 1p1h structure as illustrated in Fig.~\ref{f_wav3}(a)(black),
which shows the coefficients $c^\rho_{k'k}$ for the different 1p1h excitations.
The components of the electric octupole (or E3) decay
of the $3^-_1$ level to the $0^+$ ground state
can be obtained from the product of the $c^\rho_{k'k}$
and the corresponding reduced transition matrix elements
$\langle0^+_1\|{\rm E3}\|j_{\rho k'}j_{\rho k}^{-1};3^-\rangle$;
they are shown in Fig.~\ref{f_wav3}(b)(black).
Although
the coefficients $c^\rho_{k'k}$ themselves have alternating signs, see Fig.~\ref{f_wav3}(a)(black),
{\em all} components of the E3 transition act coherently.
The collective character of the octupole state
$3^-_{\rm c}$
follows from above-mentioned coherence with respect to E3 decay. 
As a result the total E3 reduced transition matrix element,
obtained by multiplying with the effective charges $e_\nu=0.35$ and $e_\pi=1.35$
and summing all contributions,
leads to an enhanced transition probability, $B({\rm E3};3^-_{\rm c}\rightarrow0^+_1)=36$~W.u.,
to be compared with the experimental value of 34.0(5)~W.u.~\cite{Spear83}.

The fragmentation and coherence in the E3 decay of the $3^-_{\rm c}$ state
can be understood in the context of a schematic model
that assumes degenerate single-nucleon energies
below and above the shell closures (with gaps $\Delta\epsilon_\rho$)
and a surface delta interaction (SDI) between the nucleons~\cite{Isacker20}.
For the SDI it is assumed that~\cite{Brussaard77}
(i) the interaction takes place at the surface only,
(ii) the two-body force is of extreme short range,
and (iii) the probability of finding a nucleon at the nuclear surface
is independent of the shell-model orbital in which the nucleon moves.
The SDI is a crude approximation to a realistic shell-model interaction
in terms of the isovector strengths $a_{1\rho}$ of the $\nu\nu$ and $\pi\pi$ interactions,
and the isoscalar and isovector strengths $a_0$ and $a_1$ of the $\nu\pi$ interaction,  
all with an estimated value of $(25/A)$~MeV~\cite{Brussaard77}. 
With the assumptions of the
schematic model the coefficients in Eq.~(\ref{e_oph}) can be derived analytically~\cite{note1},
\begin{equation}
\bar c^\rho_{k'k}\approx\alpha_\rho(S_\rho)^{-1/2}g^\rho_{k'k},
\label{e_can}
\end{equation}
with $S_\rho\equiv\sum_{kk'}(g^\rho_{k'k})^2$ and
\begin{equation}
g^\rho_{k'k}=
(-)^{j_{\rho k}-1/2}[j_{\rho k}][j_{\rho k'}]
\Bigl(\begin{array}{ccc}
j_{\rho k'}&j_{\rho k}&3\\\sfrac12&-\sfrac12&0
\end{array}\Bigr),
\label{e_gan}
\end{equation}
where the symbol between brackets is a Wigner $3j$ coefficient~\cite{Talmi93}
and $[j]\equiv\sqrt{2j+1}$.
For $^{208}$Pb,  $S_\nu=25.39$ and $S_\pi=16.29$.
The coefficients $\alpha_\nu$ and $\alpha_\pi$ in Eq.~(\ref{e_can}) are obtained
from the diagonalization of the $2\times2$ matrix
\begin{equation}
\left[\begin{array}{cc}
\displaystyle\Delta\epsilon_\nu-\frac{a_{1\nu}}{2}S_\nu&
\displaystyle-\frac{3a_0+a_1}{4}\sqrt{S_\nu S_\pi}\\
\displaystyle-\frac{3a_0+a_1}{4}\sqrt{S_\nu S_\pi}&
\displaystyle\Delta\epsilon_\pi-\frac{a_{1\pi}}{2}S_\pi
\end{array}\right], 
\label{e_matan}
\end{equation}
which decouples approximately from the rest of the 1p1h space.
This defines the neutron and proton octupole excitations $3^-_{{\rm c}\nu}$ and $3^-_{{\rm c}\pi}$,
which are strongly coupled
by the off-diagonal element in the matrix~(\ref{e_matan}) due to the $\nu\pi$ interaction,
giving rise to a combination
\begin{equation}
|3^-_{\rm c}\rangle=\alpha_\nu|3^-_{{\rm c}\nu}\rangle+\alpha_\pi|3^-_{{\rm c}\pi}\rangle 
\label{e_wfan}
\end{equation}
at low energy,
which is symmetric (i.e., the $\alpha_\nu$ and $\alpha_\pi$ have the same sign),
and an anti-symmetric (sometimes called isovector) combination at higher energy.

The coefficients $\bar c^\rho_{k'k}$ of Eq.~(\ref{e_can})
are shown in Fig.~\ref{f_wav3}(a)(blue).
The $3^-_{\rm c}$ states in the realistic and schematic calculation
have a similar structure with an overlap
$\sum_{\rho kk'}c^\rho_{k'k}\bar c^\rho_{k'k}\approx0.87$.
The strong mixing between $3^-_{{\rm c}\nu}$ and $3^-_{{\rm c}\pi}$, obtained with SDI,
indicates that the $\nu\pi$ interaction
is a key component of octupole collectivity (see also below).
The coherence of the E3 decay can also be proven analytically
for the schematic model, leading to~\cite{Isacker20}
\begin{equation}
B({\rm E3};3^-_{\rm c}\rightarrow0^+_1)=
\biggr(\sum_\rho e_\rho\alpha_\rho\frac{S^{(3)}_\rho}{\sqrt{S_\rho}}\biggr)^2e^2b^6,
\label{e_be3an}
\end{equation}
where $b$ is the length parameter of the harmonic oscillator
and $S^{(\lambda)}_\rho\equiv\sum_{kk'}(g^\rho_{k'k})^2I_{n_{\rho k}l_{\rho k}n_{\rho k'}l_{\rho k'}}^{(\lambda)}$,
in terms of radial integrals that are all positive.
The corresponding expression in the geometric collective model
reads (see Sect.~6.3 of Ref.~\cite{BM75})
\begin{equation}
B({\rm E3};3^-_{\rm c}\rightarrow0^+_1)=
\biggl(\frac{3}{4\pi}ZeR^3\sqrt{\frac{\hbar \omega_3}{2C_3}}\biggr)^2,
\label{e_be3BM}
\end{equation}
where $\sqrt{\hbar\omega_3/2C_3}$ is the amplitude of the octupole vibration.

A further remark concerns the origin of octupole collectivity. 
A different set of coefficients $\tilde c^\rho_{k'k}$ in Eq.~(\ref{e_oph}) is obtained
if the $\nu\pi$ component of the realistic interaction is put to zero,
see Fig.~\ref{f_wav3}(c)(red).
This leads to a strongly reduced fragmentation of the $3^-_{{\rm c}\rho}$ states:
Without $\nu\pi$ interaction octupole collectivity is greatly diminished.
It is also diminished if the $\nu\nu$ and $\pi\pi$ interactions are omitted
but the loss of collectivity is less important in that case.
Therefore, the collectivity of the octupole states
exists predominantly by virtue of the $\nu\pi$ interaction,
which couples the neutron and proton $3^-_{{\rm c}\rho}$ states
and generates their collective structure.

{\it Octupole states in odd-mass nuclei.}
We consider an odd-mass nucleus with one neutron more than a doubly magic nucleus.
The case of a proton and/or a hole coupled to a doubly magic nucleus
can be treated in a similar fashion.
The shell-model calculation is carried out in a basis with 1p0h and 2p1h states,
\begin{equation}
|j_{\nu r'}\rangle
\quad{\rm and}\quad
|(j_{\rho k'}j_{\rho k}^{-1})^{(J_k)}j_{\nu s'};J\rangle.
\label{e_2p1h}
\end{equation}
The first state represents a neutron in the orbital $j_{\nu r'}$
and in the second state the 1p1h configuration is coupled to angular momentum $J_k$,
which is subsequently coupled with $j_{\nu s'}$ to total $J$.
All possible combinations of $j_{\rho k'}$, $j_{\rho k}^{-1}$, $J_k$, and $j_{\nu s'}$
lead to an overcomplete, non-orthogonal basis,
which can be reduced to a complete, orthogonal one by diagonalizing the overlap matrix.
This basis may include spurious components with $J^\pi_k=1^-$,
which are removed by adding a center-of-mass term~\cite{Gloeckner74}
to the physical Hamiltonian~(\ref{e_ham}).
The calculation in the basis~(\ref{e_2p1h})
requires the evaluation of different types of matrix element,
which can be done by standard angular-momentum recoupling methods.

States of the odd-mass nucleus can be obtained
by coupling a particle to the octupole phonon of the core.
If such configurations occur at low excitation energy
and if they are not fragmented over many states by the nuclear interaction,
a simple `octupole' description of the odd-mass nucleus can be envisaged.
In this octupole model, besides single-particle states,
only states obtained from the coupling of a particle to the octupole phonon are considered.
A case of particular interest occurs
if two orbitals $j_{\nu r'}$ and $j_{\nu s'}$ in the basis~(\ref{e_2p1h})
have opposite parity and satisfy $|j_{\nu s'}-j_{\nu r'}|=|l_{\nu s'}-l_{\nu r'}|=3$.
In $^{209}$Pb this happens for $j_{\nu r'}=g_{9/2}$ and $j_{\nu s'}=j_{15/2}$.
In that case the single-particle state $|j_{\nu r'}\rangle$
and the octupole-particle state $|3^-_{\rm c}\times j_{\nu s'};J\rangle$ may mix if $J=j_{\nu r'}$,
giving rise to the eigenstates
\begin{align}
|J_{{\rm o}1}\rangle={}&
\alpha|j_{\nu r'}\rangle+
\beta|3^-_{\rm c}\times j_{\nu s'};J\rangle,
\nonumber\\
|J_{{\rm o}2}\rangle={}&
\beta|j_{\nu r'}\rangle-
\alpha|3^-_{\rm c}\times j_{\nu s'};J\rangle,
\label{e_states2}
\end{align}
where the subscript `o' indicates that these are states in the octupole model.
Since they are obtained from the diagonalization of a $2\times2$ Hamiltonian matrix 
in the basis $|j_{\nu r'}\rangle$ and $|3^-_{\rm c}\times j_{\nu s'};J\rangle$,
they carry an additional index 1 or 2.

In the particle-vibration coupling approach of the geometric collective model~\cite{BM75}
the diagonal elements of the $2\times2$ matrix are
\begin{align}
\langle j_{\nu r'}|\hat H|j_{\nu r'}\rangle={}&
E_0+\epsilon_{\nu r'},
\label{e_dmeo}\\
\langle3^-_{\rm c}\times j_{\nu s'};J|\hat H|3^-_{\rm c}\times j_{\nu s'};J\rangle={}&
E_0+\epsilon_{\nu s'}+E_{\rm x}(3^-_{\rm c}),
\nonumber
\end{align}
where $E_0$ is the ground-state energy of the doubly magic nucleus
and $E_{\rm x}(3^-_{\rm c})$ the excitation energy of its $3^-_{\rm c}$ state.
The mixing between the single-particle and octupole-particle states
is derived from the phonon character of the octupole excitation,
leading to (see Sect.~6.5 of Ref.~\cite{BM75})
\begin{align}
&h(j_{\nu r'},3^-_{\rm c}j_{\nu s'})\equiv
\langle j_{\nu r'}|\hat H|3^-_{\rm c}\times j_{\nu s'};J=j_{\nu r'}\rangle
\label{e_omeBM}\\
&=\varphi
\sqrt{\frac{7}{4 \pi}}[j_{\nu s'}]
\Bigl(\begin{array}{ccc}
j_{\nu s'}&j_{\nu r'}&3\\\sfrac12&-\sfrac12&0
\end{array}\Bigr)
\sqrt{\frac{\hbar \omega_3}{2C_3}}
\langle j_{\nu s'}|k_3(r)\vert j_{\nu r'}\rangle,
\nonumber
\end{align}
where $k_3(r)=R_0\partial V(r)/\partial r$ is the form factor of the octupole oscillation
and $\varphi$ is a phase.

{\it Application to $^{209}$Pb.}
We test the validity of the particle-vibration coupling approach of the geometric collective model
by comparing it to a shell-model calculation in a 1p0h+2p1h basis.
In a {\em first} step the diagonalization of the shell-model Hamiltonian~(\ref{e_ham})
for the nucleus $^{208}$Pb determines the coefficients $c^\rho_{k'k}$.
{\em Next}, we evaluate the diagonal matrix elements 
$\langle j_{\nu r'}|\hat H|j_{\nu r'}\rangle$
and
$\langle3^-_{\rm c}\times j_{\nu s'};J|\hat H|3^-_{\rm c}\times j_{\nu s'};J\rangle$.
While the first is just the expression~(\ref{e_dmeo}),
the second is calculated microscopically
with use of the coefficients $c^\rho_{k'k}$ derived for $^{208}$Pb.
The off-diagonal matrix element $h(j_{\nu r'},3^-_{\rm c}j_{\nu s'})$ in Eq.~(\ref{e_omeBM})
is also calculated microscopically with the expression
\begin{align}
&-\sqrt{\frac{7}{2J+1}}
\biggl(\sum_{k'k}c^\nu_{k'k}
\langle j_{\nu r'}j_{\nu s'}^{-1};3|\hat V_{\nu\nu}|j_{\nu k'}j_{\nu k}^{-1};3\rangle
\nonumber\\&+
\sum_{k'k}c^{\pi}_{k'k}
\langle j_{\nu r'}j_{\nu s'}^{-1};3|\hat V_{\nu\pi}|j_{\pi k'}j_{\pi k}^{-1};3\rangle\biggr).
\label{e_ome}
\end{align}
Although the individual terms in the sums in Eq.~(\ref{e_ome}) are small,
they all act coherently to yield a large off-diagonal matrix element~\cite{Ralet19}.

The coefficients $\alpha$ and $\beta$ in Eq.~(\ref{e_states2})
are obtained by diagonalizing a $2\times2$ Hamiltonian matrix
with elements calculated in the shell model.
In a {\em final} step a diagonalization of the Hamiltonian~(\ref{e_ham})
is carried out in the full 1p0h+2p1h basis,
followed by the calculation of the overlaps $|\langle J_k^\pi|J_{{\rm o}i}^\pi\rangle|^2$,
where $|J_k^\pi\rangle$ is the $k^{\rm th}$ eigenstate in the full space
and $|J_{{\rm o}i}^\pi\rangle$, $i=1,2$, an eigenstate in the two-dimensional space.
The states~(\ref{e_states2}) constitute a good approximation
to the lowest shell-model states in the 1p0h+2p1h space
if $|\langle J_k^\pi|J_{{\rm o}i}^\pi\rangle|^2\lesssim1$ for $k=i=1,2$,
and $|\langle J_k^\pi|J_{{\rm o}i}^\pi\rangle|^2\approx0$ otherwise.

\begin{figure}
\includegraphics[width=8.5cm]{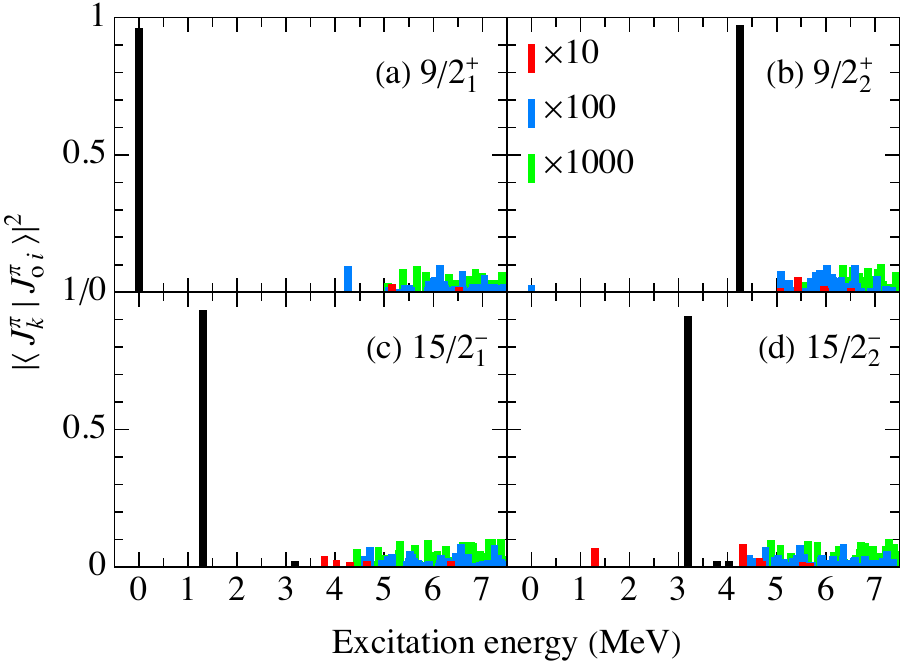}
\caption{
Overlaps $|\langle J_k^\pi|J_{{\rm o}i}^\pi\rangle|^2$ (see text)
for the $J^\pi=9/2^+$ and $15/2^-$ states in $^{209}$Pb.
The two eigenstates $|J_{{\rm o}i}^\pi\rangle$, $i=1,2$,
are obtained in the octupole model
with the single-particle orbitals $g_{9/2}$ and $j_{15/2}$,
and their coupling to $3^-_{\rm c}$.
The eigenstates $|J_k^\pi\rangle$ are obtained
with the realistic shell-model Hamiltonian in the 1p0h+2p1h space.
For display purposes the squared overlaps
are multiplied by 1, 10, 100, or 1000, indicated in black, red, blue, or green, respectively.}
\label{f_wav915} 
\end{figure}
Figure~\ref{f_wav915} provides an illustration of this test.
The two-component wave functions originate
from the single-particle orbitals $g_{9/2}$ and $j_{15/2}$,
and their coupling to $3^-_{\rm c}$;
the wave functions in the 1p0h+2p1h space
have 467 and 530 components for $J^\pi=9/2^+$ and $15/2^-$, respectively.
The lower-energy eigenstate of the $2\times2$ matrix
is predominantly of single-particle character,
either $g_{9/2}$ or $j_{15/2}$,
and this finding is confirmed by the calculation in the 1p0h+2p1h space.
The upper-energy eigenstate is predominantly of octupole-particle character,
and the square of its overlap with the full-space wave function
is 0.97 (0.91) for $J_2^\pi=9/2_2^+$ ($15/2_2^-$).

\begin{figure}
\includegraphics[width=8.5cm]{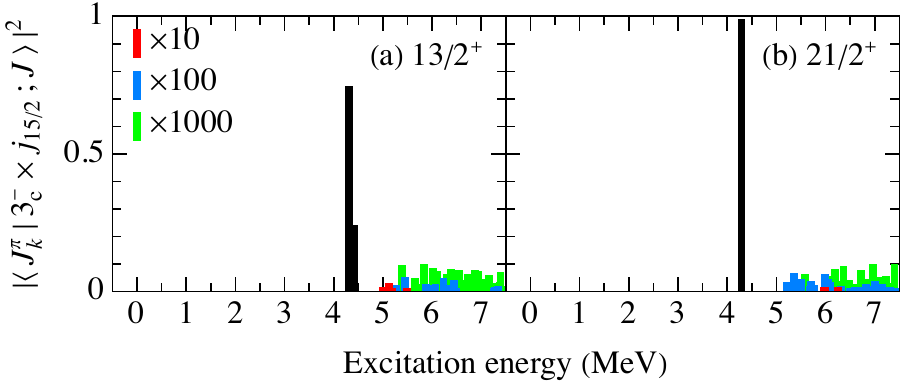}
\caption{
Distribution of a neutron in the $j_{15/2}$ orbital
coupled to the octupole excitation $3^-_{\rm c}$ of $^{208}$Pb.
The bars (in the color code of Fig.~\ref{f_wav915})
indicate the overlaps $|\langle J^\pi_k|3^-_{\rm c}\times j_{15/2};J\rangle|^2$
for $J^\pi=13/2^+,21/2^+$ in $^{209}$Pb,
where $|J^\pi_k\rangle$ are eigenstates of the shell-model Hamiltonian in the 1p0h+2p1h space.}
\label{f_wavj15_1321} 
\end{figure}
Many spin-parities are not among those available to a single neutron in the 126-184 orbitals
but can be obtained by coupling the neutron to the octupole phonon.
Are such particle-octupole states found in the low-energy eigenspectrum
of the realistic shell-model Hamiltonian?
Figure~\ref{f_wavj15_1321} shows two representative cases,
namely the states $|3^-_{\rm c}\times j_{15/2};J^\pi\rangle$ with $J^\pi=13/2^+$ and $21/2^+$,
and their distribution over the eigenstates in $^{209}$Pb.
The $J^\pi=21/2^+_1$ state is overwhelmingly (the squared overlap is 0.996)
that of a $j_{15/2}$ neutron coupled to $3^-_{\rm c}$.
In contrast, the $J^\pi=13/2^+_1$ state,
while still dominantly $3^-_{\rm c}\times j_{15/2}$ (the squared overlap is 0.74),
interacts with $|(g_{9/2}g_{9/2})^{(0)}i_{13/2}^{-1};13/2\rangle$,
which happens to be very close in energy.

The conclusion of a systematic study of all possible cases involving single-particle
as well as octupole-particle excitations in $^{209}$Pb
is that no member of the $3^-_{\rm c}\times g_{9/2}$ multiplet
shows any fragmentation in the shell-model calculation,
that the $3^-_{\rm c}\times j_{15/2}$ multiplet is pure except for $J^\pi=13/2^+$,
and that the high-spin members ($J^\pi\gtrsim11/2^-$)
of the $3^-_{\rm c}\times i_{11/2}$ multiplet are also unfragmented.
Other multiplets occur at higher energies in the region of higher level density
and display considerable fragmentation.
However, close to the full strength of particle-octupole coupled states
can be found within $\sim400$~keV around the unperturbed particle-octupole energy.
Our results demonstrate that the $3^-_{\rm c}$ state
preserves its bosonic identity while coupling with a single particle or hole. 

In the schematic model the following expression is obtained for the off-diagonal matrix element~(\ref{e_ome}):
\begin{align}
&h(j_{\nu r'},3^-_{\rm c}j_{\nu s'})
\label{e_omean}\\&\quad\approx
(-)^{j_{\nu r'}+1/2}\sqrt{7}[j_{\nu s'}]
\Bigl(\begin{array}{ccc}
j_{\nu s'}&j_{\nu r'}&3\\\sfrac12&-\sfrac12&0
\end{array}\Bigr)
f(a_{1\nu},a_0,a_1),
\nonumber
\end{align}
with
\begin{equation}
f(a_{1\nu},a_0,a_1)=
\biggl(\frac{a_{1\nu}}{2}\alpha_\nu\sqrt{S_\nu}+
\frac{3a_0+a_1}{4}\alpha_\pi\sqrt{S_\pi}\biggr).
\label{e_function}
\end{equation}
The analytic expressions~(\ref{e_be3an}) and~(\ref{e_omean})
together with the corresponding equations~(\ref{e_be3BM})
and~(\ref{e_omeBM}) in the geometric collective model,
lead to the following result:
\begin{equation}
|\langle j_{\nu s'}|k_3(r)\vert j_{\nu r'}\rangle|\approx
\frac{15}{16}\sqrt{\frac{15}{2\pi}}Z\sqrt{A}
\frac{f(a_{1\nu},a_0,a_1)}{S},
\label{e_formfactor}
\end{equation}
where $S=\sum_\rho e_\rho\alpha_\rho S^{(3)}_\rho S_\rho^{-1/2}$
and use is made of Eq.~(2.36) of Ref.~\cite{Brussaard77} for the oscillator length $b$.
Equation~(\ref{e_formfactor}) gives a simple expression for the matrix element of $R_0\partial V(r)/\partial r$
in terms of the orbitals included in the shell-model space
and of the strength parameters of the SDI.

In summary, octupole excitations in doubly magic nuclei exhibit universal symmetry properties
that explain their collective structure and phonon-like behavior.
In $^{208}$Pb the collectivity of the particle-hole wave function of the $3_1^-$ state
is mainly due to the neutron-proton interaction.
The enhanced reduced transition probability to the ground state,
$B({\rm E3};3^-_1\rightarrow0^+_1)$,
results from the fully coherent action of all participating particle-hole excitations
and is a consequence of the attractive nature of the residual interaction.
When coupled to a low-energy single particle or hole excitation,
the collective nature and bosonic identity of the octupole state is preserved:
E3 collectivity in odd-mass nuclei remains concentrated in a single state
and is not fragmented over many shell-model eigenstates. 
In the geometric collective model
the coupling of octupole degrees of freedom to a nucleon
arises from the variation in the average nuclear potential due to the collective vibration.
In the shell model this coupling results from the coherent action
of all the components of the collective state.
The findings obtained here with a realistic shell-model interaction both for $^{208}$Pb and $^{209}$Pb
agree with the geometric collective model
and confirm the phononic behavior of octupole excitations in nuclei.



\begin{thebibliography}{[9]}
\bibitem{note1}
This expression is derived in the phase convention of radial oscillator wave functions
that are positive for $r\rightarrow\infty$.
\end{thebibliography}
\end{document}